\documentstyle[psfig]{mn2e}

\title[]{GRS 1915+105: The brightest Galactic black hole}

\author[C.~Done, G.~Wardzi\'nski, and M.~Gierli\'nski] {Chris Done$^1$, Grzegorz
Wardzi\'nski$^{1,2}$ and Marek~Gierli\'nski$^{1,3}$\\
$^1$Department of Physics, University of Durham, South Road, Durham DH1 3LE,
UK\\
$^2$ Centrum Astronomiczne im.\ M. Kopernika, Bartycka 18, 00-716 Warszawa,
Poland\\
$^3$Obserwatorium Astronomiczne Uniwersytetu Jagiello{\'n}skiego, 30-244
Krak{\'o}w, Orla 171, Poland}

\date{Submitted to MNRAS}
\pagerange{\pageref{firstpage}--\pageref{lastpage}}
\pubyear{2003}

\begin{document}

\def\lh{$\ell_{\rm h}$}
\def\lth{$\ell_{\rm th}$}
\def\lnth{$\ell_{\rm nth}$}
\def\ls{$\ell_{\rm s}$}
\def\lhs{$\ell_{\rm h}/\ell_{\rm s}$}
\def\lnh{$\ell_{\rm nth}/\ell_{\rm h}$}
\def\grs{GRS~1915+105}

\def\aap{A\&A}
\def\apj{ApJ}
\def\apjl{ApJ}
\def\mnras{MNRAS}

\newcommand{\xte}{{\it RXTE}}

\topmargin = -0.5cm

\maketitle

\label{firstpage}

\begin{abstract}

We compare the evolution of spectral shape with luminosity in \grs\
with that of `normal' black holes. The pathological variability of
\grs, which probably indicates a disc instability, does not require
that \grs\ belongs in a different class to all the other objects. At
comparable fractions of Eddington luminosity its spectra and (more
importantly) apparent disc stability, are both similar to that seen in
the `normal' black holes. Its unique limit-cycle variability only
appears when it radiates at uniquely high (super-Eddington)
luminosities.

\end{abstract}


\section{Introduction}

\grs\ is a spectacularly variable accreting black hole in our
Galaxy. It was first detected in X-rays by the WATCH all sky
monitor on board the {\it GRANAT\/} satellite in 1992
(Castro-Tirado, Brandt \& Lund 1992), but achieved fame when radio
observations showed it to be the first galactic superluminal
source, implying relativistic plasma ejection (Mirabel \&
Rodr{\'{\i}}guez 1994). Several more superluminal jet sources have
since been found (GRO J1655--40, XTE J1748--288, V4641 Sgr), while
a much larger number of black holes (the {\em microquasars}) show
jet-like radio morphology (e.g. Mirabel \& Rodr{\'{\i}}guez 1999).
Mildly relativistic jets are now recognized as a common feature of
black holes at high mass accretion rates (Fender \& Kuulkers
2001).

However, \grs\ is still unique in its variability properties. On long
timescales, its outburst duration of over 10 years is unprecedented,
and completely at odds with the predictions of the disc instability
models which can successfully describe some aspects of the other
black hole transient outbursts (King \& Ritter 1998; Dubus, Hameury
\& Lasota 2001; Esin, Lasota \& Hynes 2000). On short timescales, the
X-ray variability is even more singular, showing episodes where it
continually switches between states in a quasi--regular way (Greiner,
Morgan \& Remillard 1996; Chen, Swank \& Taam 1997). Belloni et al.
(1997a) showed that these rapid changes were associated with the
accretion disc spectrum switching from hot and bright, implying a
small inner disc radius, to cooler and dimmer, with a larger
inferred radius. They interpreted this as the result of a limit-cycle
instability in the inner accretion disc, such that it is continually
emptying and refilling.

Following Belloni et al. (1997a, b), most models for the origin of
this unique limit-cycle behaviour of the accretion disc have
concentrated on the radiation pressure instability in a standard
Shakura-Sunyaev accretion disc (Shakura \& Sunyaev 1973). In the
$\alpha$ disc prescription the viscous heating is proportional to the
total pressure $P_{\rm tot}=P_{\rm gas}+P_{\rm rad}$, i.e. the sum of
the gas and radiation pressures. Where radiation pressure dominates
the heating rate is dramatically sensitive to temperature ($P_{\rm
rad}\propto T^4$ compared to $P_{\rm gas}\propto T$), and the
radiative and convective cooling in the disc cannot keep pace. By
itself, this would just lead to the classic thermal-viscous
instability (Lightman \& Eardley 1974; Shakura \& Sunyaev 1976).
However, there is another stable disc solution at high temperatures,
where optically thick advective cooling becomes important (Abramowicz
et al.~1988). Putting all this together at a given radius and $\alpha$
gives rise to an S-curve on a plot of disc surface density, $\Sigma$
versus mass accretion rate through the disc, $\dot{m}\propto T^4$. The
lower, middle and upper branches of the S corresponding to heating
$\propto P_{\rm gas}$ balanced by radiative cooling (stable), heating
$\propto P_{\rm rad}$ balanced by radiative cooling (unstable) and
heating $\propto P_{\rm rad}$ balanced by advective cooling (stable;
Abramowicz et al.~1988; Chen et al. 1995). No steady state equilibrium
solution is possible at mass accretion rates corresponding to the
middle branch. Instead there is a continual limit-cycle with the
material switching between the two stable states, one which has too
high a mass accretion rate, leading to the emptying of the disc, and
one with too low as mass accretion rate, so the disc refills.

However, it is difficult to understand why the radiation pressure
instability is  not apparent in other black hole X-ray transients (see e.g.
Gierli{\'n}ski \& Done 2003b, hereafter GD03). The only features
distinguishing black holes are mass and spin. \grs\ is the most
massive galactic black hole known at $\sim$14~M$_\odot$
(Greiner, Cuby \& McCaughrean 2001a), but this is not
significantly larger than the typical $10 M_\odot$ of the other
black holes (Bailyn et al.~1998). It may well be spinning close
to maximal, but so may GRO~J1655--40 (Zhang, Cui \& Chen 1997;
Cui, Zhang \& Chen 1998). If the difference is not in the black
hole itself, then it must be connected to the accretion flow.
\grs\  accretes at high fractions of the Eddington limit, but so
do many other black hole transients (see e.g the review by
Tanaka \& Lewin 1995). \grs\ has a jet, which is clearly linked
to limit-cycle behaviour of the disc (Pooley \& Fender 1997;
Eikenberry et al. 1998), but the ratio of radio to X-ray power
is similar to that of other X-ray transients (Fender \& Kuulkers
2001). Why then is the variability from \grs\ unique?

Recently, Done \& Gierli{\'n}ski (2003; hereafter DG03) used the
huge {\it Rossi X-ray Timing Explorer} (\xte\/) database to
systematically analyze the spectra of many black hole systems.
These were all consistent with the {\em same} spectral evolution
as a function of Eddington luminosity\footnote{We use Eddington
luminosity $L_{\rm Edd} = 1.26\times10^{38} M/$M$_\odot$ erg
s$^{-1}$ for a mass $M$ throughout this paper.} fraction,
$L/L_{\rm Edd}$, so these form a `normal' black hole sample for
comparison with \grs. Here we analyze the {\it RXTE\/} data on
\grs\ and plot it together with the `normal' black holes. We show
that \grs\ is unique in being the only black hole binary which
spends any considerable time above $L/L_{\rm Edd} \sim 1$, and
speculate that this is the trigger for the limit-cycle variability
in \grs.

The reason \grs\ alone gets to such consistently high mass accretion
rates can be simply related to its evolutionary state. The secondary
is a giant (Greiner et al.~2001b), so Roche lobe overflow occurs in a
much wider binary than for a main sequence star. The 33.5-day period
of \grs\ is by far the longest of any low-mass X-ray binary (LMXB)
known. This implies a huge disc size, as the outer radius is set by
tidal truncation. During quiescence the disc builds up an immense
reservoir of mass, which can fall onto the black hole when the
outburst is triggered (King \& Ritter 1998). The underlying cause of
all the unique long and short-term variability of \grs\ is then
ultimately linked to the evolution of the huge disc structure, which
can contain enough material to maintain super-Eddington accretion
rates over timescales of tens of years.

\section{Data selection and reduction}


\begin{figure*}
\centerline{{\psfig{file=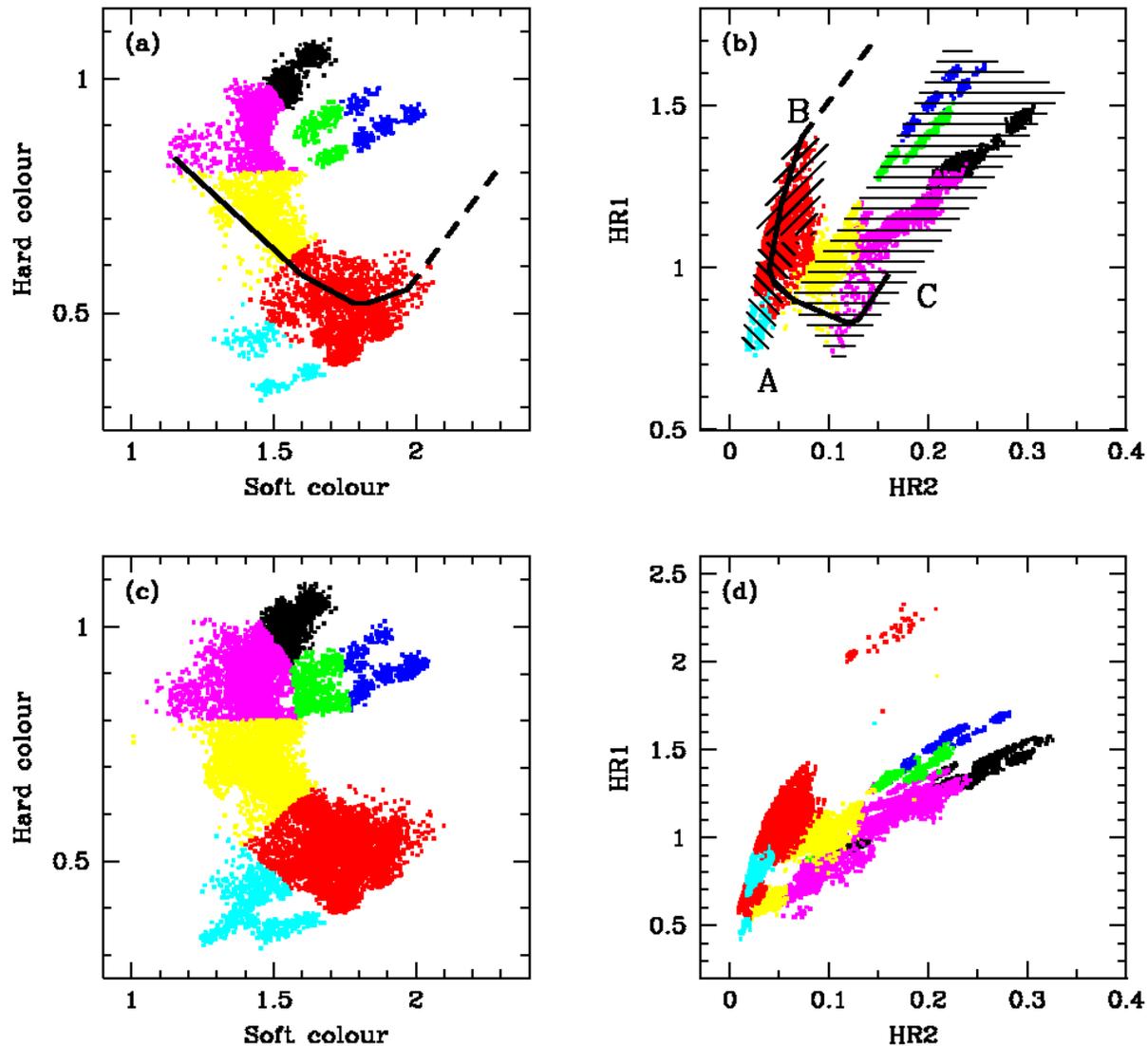,width=0.9\textwidth}}}

\caption{(a) shows the intrinsic colour-colour diagram for each
128-s spectrum of \grs\, using the data selection of B00. These
colours are corrected for interstellar absorption and for the instrument
response, giving a measure of the intrinsic spectral slope in
the 3--6.7~keV (soft colour) and 6.7--16.0~keV (hard colour)
bandpass (see DG03). (b) shows the corresponding instrument
colour-colour diagram as used by B00. Regions marked in the same
colours in panels (a) and (b) correspond to the same data. The
shading indicates the spectral states A, B and C defined by B00,
while the solid/dotted line is again from B00, indicating the
path of a typical instability. The B00 data are on 1-s
resolution, and so their colours can follow the very rapid
variability in the flares (indicated by the dotted line) which
are not sampled by the 128-s spectra used here. Panels (c) and
(d) reproduce panels (a) and (b) but for all observations from
PCA epochs 2--4. The intrinsic diagram looks very similar, but
the instrument diagram is very different due to gain changes in
the detector. There is no longer a unique correspondence between
the intrinsic spectral shape and instrument count ratios (e.g.
the red region indicating a particular spectral shape is split
into three separate regions in instrument counts).  }

\label{fig:bell_col}
\end{figure*}


\xte\/ has been in operation since January 1996. The Proportional
Counter Array (PCA) on board \xte\/ had undergone several major
gain changes, marking five instrument epochs. Here we analyze all
PCA data before the end of PCA epoch 4, i.e. observations between
1996 April 6 and 2000 May 11 (epoch 5 is marked by the loss of the
propane layer for detector 0). We use {\sc ftools} 5.2 and follow
standard data reduction procedure as recommended by the {\it
RXTE\/} team. To maximize the number of observations which can be
used we select data from PCA detector 0 only, as it is switched on
most of the time. The source is dramatically variable so we
extract data in 128-s bins, giving a quite overwhelming number of
16489 individual spectra. There is variability on shorter
timescales (see e.g. Belloni et al.~2000; hereafter B00) but this
is the shortest timescale on which the spectra are generally
limited by systematic errors (set to 1 per cent) rather than
statistical errors below 10 keV. We also accumulate both light
curves and spectra for each of the 500 separate pointings,
designated by unique observation IDs, typically each giving a few
kiloseconds of data. The light curves for each observation ID are
extracted over 3--20 keV with the 16-second time resolution of the
Standard 2 data.

\section{Spectral modelling}

We use {\sc xspec} (Arnaud 1996) to fit the PCA data in 3--20~keV
range with 1 per cent error added in each energy channel to represent
the systematic uncertainties of the detector response. We assume the
abundances of Anders \& Ebihara (1982) unless stated otherwise, and
fix the interstellar galactic absorbing column at $4.7\times 10^{22}$
cm$^{-2}$ (Chaty et al.~1996). For calculating luminosities we assume
distance of 12.5~kpc (Mirabel \& Rodr{\'{\i}}guez 1994 but see Fender
et al. 1999). All parameter ranges are quoted for $\Delta \chi^2$ =
2.7.

Following DG03 we fit a physically motivated continuum model
consisting of a multicolour disc spectrum which also forms the
seed photons for Compton scattering to higher energies. We use the
{\sc diskbb} model (Mistuda et al.~1984) to describe the disc
emission. The temperature is constrained to $kT_{\rm disc} \ge
0.4$~keV. so as not to produce an unphysically large component
outside of the energy range of the PCA data. For Comptonization we
use {\sc thcomp} (not included in the standard distribution of
{\sc xspec}, see Zdziarski, Johnson \& Magdziarz 1996),
parameterized by asymptotic photon spectral index ($\Gamma \ge 1$)
and electron temperature ($kT_{\rm comp} \le$ 100 keV).

\subsection{Simple model}

The physically motivated continuum model does not fully describe
the observed spectra from X-ray binaries as there is also Compton
reflection from the disc. Following DG03 we approximate this by
adding a broad Gaussian line (width fixed at $0.5$~keV) and
smeared edge (index for photoelectric cross-section fixed at -2.67
and smearing width of 7~keV; see Ebisawa 1991). This model
spectrum is described as M1 in Table~\ref{tab:models} and
throughout the text.

We use this to fit each 128-s spectrum from \grs. It gives a
good description of most of the data; only 5 per cent have
reduced $\chi^2/\nu > 1.5$ (where $\nu$ is number of degrees of
freedom). We follow the approach of DG03 in compressing all the
resulting spectral information from the model fits into two {\em
intrinsic} colours, which are calculated by integrating the
(absorption corrected) model flux over 4 energy bands (3--4,
4--6.4, 6.4--9.7, 9.7--16~keV) to form soft and hard colours.
These roughly describe the unabsorbed spectral slope from 3--6.4
keV and 6.4--16 keV, respectively.

This method contrasts with that of the more usual instrument
colours, which are defined using the ratios of observed counts in
given energy bands (as opposed to modelled flux ratios). Both
methods have their advantages and disadvantages.  The advantage of
our method is that modelling corrects for both instrument response
and absorption in a single step, and gives colours which reflect
the underlying source spectrum. Thus many objects, with data taken
from many different instruments and/or gain epochs, with very
different absorption columns, can be easily plotted together on
the same diagram. Instrument colours can be modified to correct
for different gain epochs (e.g. Di Salvo et al. 2003, van Straaten
et al. 2003), but different absorptions prevent a straightforward
comparison of many different objects. Another advantage of our
approach is that the model spectra allow us to estimate the {\em
bolometric} luminosity, which is surely a major physical driver of
the spectral evolution. The disadvantage of course is that these
luminosities are model dependent (though the colours are not as
long as absorption does not substantially affect the spectrum
above 3 keV: DG03). More importantly, the requirement for good
signal-to-noise spectra for the model fitting means that the
intrinsic colours have lower time resolution than the instrument
ones.

Belloni et al. (2000: hereafter B00) used instrument colours for
their major study of the variability of \grs\ as seen in Epoch 3.
Fig.~\ref{fig:bell_col}a and b shows a comparison of our derived
{\em intrinsic} colours (defined by model flux ratios) for the
Epoch 3 data with those of the {\em instrument} colours (defined
by observed count ratios) as used by B00. Data from different
regions of Fig.~\ref{fig:bell_col}a are colour-coded, so that
their corresponding position on Fig.\ref{fig:bell_col}b can easily
be identified, so that we can use the B00 insights into the nature
of the variability as a function of colour. In particular, they
unified the incredibly rich variability patterns into transitions
between 3 main spectral states -- A, B and C which they associate
with a disc instability (Belloni et al.~1997a, b). These states
are shown by the shaded areas in Fig.~\ref{fig:bell_col}b, while
the thick black line shows the track in colour-colour space of a
dramatic transition ($\lambda$-type variability, taken from
observation ID 10408-01-38-00 as shown in figure 2m of B00). The
dotted line shows the continuation of this track which is seen in
high time resolution data. B00 are able to get a good instrument
colour on 1-s time resolution, as opposed to the 128-s spectra
needed for the model fitting used here.

While we cannot follow the most rapid variability,
Figs.~\ref{fig:bell_col}c and \ref{fig:bell_col}d show some of the
power of our {\em intrinsic} colour approach when combining all the
data from Epochs 2--4. The intrinsic colour-colour diagram is very
similar to that derived just from B00 data (compare panels a and c in
Fig.~\ref{fig:bell_col}), whereas the straightforward instrument
colours are strongly affected, with data from different Epochs ending
up at very different position in the diagram (Epoch 2 has much higher
HR1 and HR2 than epoch 3, while Epoch 4 has much lower HR1 and HR2).
While these gain change effects can be corrected for (di Salvo et
al. 2003; van Straaten et al. 2003), the key issue in this paper is
that the unabsorbed flux ratios shown in Fig.~\ref{fig:bell_col}c can
be {\em directly compared} with the unabsorbed colours from many other
black holes given in e.g. DG03. We can use this straightforward
comparison of multiple objects to search for the differences between
\grs\ and the `normal' black holes which underlay its unique
variability.

\subsection{Variability as a function of colour}
\label{sec:variability}

B00 show that the unique limit-cycle variability of \grs\ occurs
in only $\sim$50 per cent of the data. Rather than do a detailed
classification of the variability pattern (as in B00), we simply
parameterize the variability in a given observation (one
individual {\it RXTE\/} pointing with unique observation ID) by
calculating the fractional r.m.s. (variance divided by mean
intensity) from its light curve with resolution of 16 s. Some of
the observations are too short for this to be reliable, so we
restrict this to light curves in which there are more than 128
data points (i.e.\ over 2048 s of data).  This fractional r.m.s.
represents the integral of the power spectrum from frequencies
lower than 1/2048 to 1/32 Hz. The low frequency power spectrum is
generally flatter than $f^{-1}$ (e.g. Morgan et al. 1997) so the
fractional r.m.s. is not strongly dependent on the length of the
light curve.

Assuming that the power spectrum is stable for each type of
variability, we can use the individual observation IDs identified
by B00 as being characteristic of each class to get an estimate of
their fractional r.m.s.. The state transition classes $\mu,
\theta, \lambda, \kappa, \rho, \nu, \alpha, \beta, \delta$ and
$\gamma$ have $17, 36, 52, 54, 35, 47, 27, 49, 16$ and $7$ per
cent variability, respectively, while the single state classes
$\phi$ and $\chi$ have $7$ and $3$ per cent, respectively. While
it is plain that our crude fractional r.m.s. criteria cannot make
subtle distinctions between the types of state transition,
nevertheless it is clear that restricting the fractional r.m.s. to
less than 5 per cent will select typically only the $\chi$ stable
state (through may include a little contamination by the $\phi$
and $\gamma$ type of variability), while selecting fractional
r.m.s. greater than 40 per cent will pick out the dramatic state
transition variability of type $\lambda, \kappa, \nu, \beta$
(perhaps with a little contribution from $\theta$ and $\rho$
states. It is these dramatic variability patterns which are unique
to \grs\ which we want to identify.

\begin{figure}
\begin{tabular}{c}
{\psfig{file=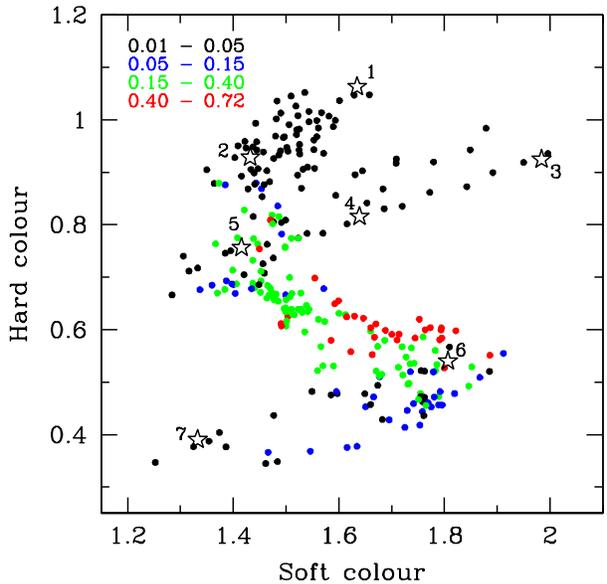,width=0.45\textwidth}}
\end{tabular}

\caption{The  intrinsic colour-colour diagram of the individual PCA
pointings with $>2048$ s of data. These cover the full range of
spectral behaviour seen in Fig.~\ref{fig:bell_col}c. The points are
colour coded according to the amount of fractional r.m.s. variability present
in the 3--20 keV light curve for each observation as
given in the key in the top-left corner of the figure. The numbered stars
show the positions of the individual spectra used for spectral
analysis in Sec.~\ref{sec:ObsIDs}.}

\label{fig:rms}
\end{figure}

The intrinsic colours for the resulting 292 spectra were found using
the model described above. Obviously, some of the spectra are averaged
over periods of high variability, including the limit-cycle spectral
transitions. These do {\em not} represent real physical spectra: the
average of two blackbodies of different temperatures is {\em not\/} a
blackbody at the average temperature, nor is the average of two power
laws of different spectral slopes given by a single power law of the
average slope!  Instead, the colours extracted from these spectra give
the time averaged position of the source on the colour-colour diagram.
Fig.~\ref{fig:rms} shows these intrinsic colours, with r.m.s.
variability indicated.  This illustrates the result of B00 that the
limit-cycle variability which is unique to \grs\ is associated with
very particular spectral shapes which mark out the track of the
instability.


\subsection{Detailed analysis of selected spectra}
\label{sec:ObsIDs}


\begin{table}
\begin{tabular}{ccccc}
\hline

Obs. & Obs. ID & Live time & Count rate & r. m. s. \\
& & (s) & (s$^{-1}$) & per cent \\
\hline
1 & 30703-01-34-00 &  2400  & 873$\pm$2  & 4.0$\pm$0.3 \\
2 & 20402-01-29-00 &  4208  & 1157$\pm$3 & 2.9$\pm$0.2 \\
3 & 10408-01-27-00 &  7552  & 1562$\pm$3 & 1.9$\pm$0.1 \\
4 & 30402-01-09-00 &  3712  & 1751$\pm$4 & 1.5$\pm$0.1 \\
5 & 30184-01-01-00 & 15184  & 1896$\pm$4 & 1.3$\pm$0.1 \\
6 & 20402-01-55-00 &  7552  & 3267$\pm$8 & 3.9$\pm$0.2 \\
7 & 30703-01-08-00 &  4000  & 1099$\pm$3 & 3.5$\pm$0.2 \\
\hline

\end{tabular}

\caption{Log of seven PCA observations selected for detailed
spectral analysis. The root mean square variability is calculated from
3--20 keV light curves with 16-s resolution.}

\label{tab:Obs IDs}
\end{table}



\begin{table*}
\begin{tabular}{clp{9cm}}
\hline
Model & {\sc xspec} components & Description \\
\hline

M1 & {\sc wabs*smedge*(diskbb+thcomp+gaussian)} & Simple model
of multicolour disc with thermal Comptonization. Reflection
features are approximated by broad Gaussian and smeared edge.
Absorption with solar abundances fixed at $N_H =
4.7\times10^{22}$ cm$^{-2}$.\\\\

M2 & {\sc wabs*(diskbb+thcomp+relrep)} & Multicolour disc,
thermal Comptonization and its ionized Compton reflection,
relativistically smeared. Absorption with solar abundances fixed
at $N_H = 4.7\times10^{22}$ cm$^{-2}$. \\\\

M3 & {\sc varabs*(diskbb+thcomp+relrep)} & The same as M2, but
with anomalous abundance absorption of Si and Fe. \\\\

M4 & {\sc varabs*(diskbb+thcomp+relrep+6$\times$gaussian)} & The
same as M3, but with narrow resonant absorption Lyman $\alpha$,
$\beta$ and $\gamma$ lines from He-like and H-like iron. The
absorption lines are assumed to be narrow and the $\beta/\alpha$
and $\gamma/\alpha$ ratios are fixed  to the oscillator strength
ratios. This is our best physical model. \\\\

M5 & {\sc varabs*smedge*(diskbb+thcomp+gaussian)} & The same as
M1, but with anomalous abundance absorption of Si and Fe. This
is a very fast model which gives very similar colours to M4.\\

\hline

\end{tabular}

\caption{Summary of spectral models used in this paper.}

\label{tab:models}
\end{table*}



\begin{table*}
\begin{tabular}{ccccccc}
\hline

Obs. & $kT_{\rm disc}$ & $\Gamma$ & $kT_{\rm comp}$ & $L_{\rm disc}$ & $L_{\rm comp}$ & $\chi^2/\nu$ \\
& (keV) & & (keV) & \multicolumn{2}{c}{($10^{38}$ erg s$^{-1}$)} \\
\hline

1 & $ < 0.74$  & $2.00_{-0.09}^{+0.05}$ & $>10$ & $2.6\pm1.6$ & $5.1_{-1.3}^{+2.3}$ & 16.4/36\\
2 & $0.82_{-0.19}^{+0.15}$ & $2.15\pm 0.16$ & $> 7$ & $3.5\pm0.4$ & $4.2_{-0.7}^{+1.5}$ & 10.5/36\\
3 & $1.36_{-0.63}^{+0.60}$ & $1.97_{-0.85}^{+0.18}$ & $5.2_{-1.4}^{+1.1}$ & $< 6.3$ & $4.2_{-2.0}^{+4.1}$ & 10.2/36\\
4 & $0.79_{-0.38}^{+0.22}$ & $2.16_{-0.18}^{+0.10}$ & $5.0_{-0.7}^{+0.8}$ & $< 4.8$ & $5.8_{-1.2}^{+3.9}$ & 18.7/36\\
5 & $< 0.87$ & $2.35_{-0.14}^{+0.11}$ & $5.5_{-0.7}^{+1.1}$ & $ <6.4 $ & $6.4_{-1.2}^{+5.1}$ & 10.8/36\\
6 & $0.90^{+0.13}_{-0.05}$ & $2.70_{-0.15}^{+0.16}$ & $3.2\pm 0.2$ & $< 3.5$ & $11.1_{-1.7}^{+1.8}$ & 5.1/36\\
7 & $1.17_{-0.01}^{+0.02}$ & $< 1.26$ & $2.14_{-0.02}^{+0.06}$ & $6.64\pm0.05$ & $0.98\pm0.01$ & 62.2/36\\
\hline

\end{tabular}

\caption{Best-fitting parameters for the selected spectra using
the standard model M1 (see Table~\ref{tab:models}).
Luminosities have been calculated for the distance of 12.5 kpc
and the disc inclination of 66$^\circ$. Eddington luminosity for
a 14~M$_\odot$ black hole is $17.6\times10^{38}$ erg s$^{-1}$.}

\label{tab:simple}
\end{table*}

\begin{figure}
\centerline{{\psfig{file=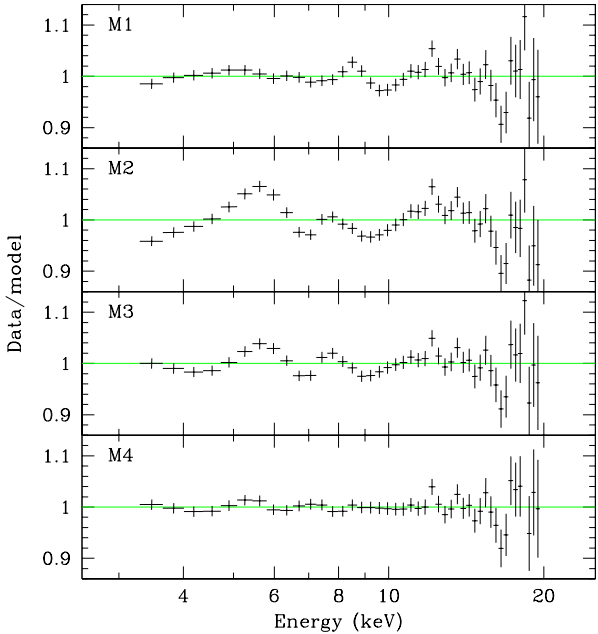,width=0.45\textwidth}}}

\caption{Residuals (data/model) for spectrum 7
(Table~\ref{tab:Obs IDs}) fit with different models as summarized in
Table~\ref{tab:models}. M4 is the most complex, but is the only
physically motivated model which can fit this spectrum.}

\label{fig:residuals}
\end{figure}

We use Fig.~\ref{fig:rms} to select seven observations which do
not show significant variability and cover full range of
behaviour in the colour-colour diagram. The chosen points are
marked with numbered stars on Fig.~\ref{fig:rms}, and listed in
Table \ref{tab:Obs IDs}. We will refer to these spectra as
S1--S7. The best-fitting parameters of the simple model M1 are
shown in Table~\ref{tab:simple}. We also show unabsorbed
luminosities of the disc ($L_{\rm disc}$) and Comptonized
($L_{\rm comp}$) components, calculated for a distance of 12.5
kpc and disc inclination of 66$^\circ$. The disc is not
significantly detected for S3--S6. The fits are generally an
excellent description of the data, despite their use of a
phenomenological broad line and smeared edge to approximately
model the reflection features. The one exception to this is S7,
which has $\chi^2/\nu=62.2/36$. Fig.~\ref{fig:residuals} (panel
M1) shows the residuals of this fit in terms of the ratio of
data/model.


In the next step we replace the phenomenological description of
the reflection features with the reflection code of {\.Z}ycki,
Done \& Smith (1998). This calculates the self-consistent line
emission and reflected continuum from an ionized,
relativistically smeared disc illuminated by a Comptonized
continuum. We fix the iron abundance at 3 times solar (Lee et
al.~2002). The model is parameterized by amplitude of reflection
($0 \le \Omega/2\pi \le 2$), reflector ionization, $\xi$, and
inner disc radius ($6 \le R_{\rm in}/R_g \le 1000$, where $R_g
\equiv GM/c^2$). We call this model M2 (see
Table~\ref{tab:models}). Fitting results in
Table~\ref{tab:reflection} show that the fit is generally
adequate except for S7 again, but the $\chi^2$ is always {\em
worse} than for the simple model M1 fits. The physical
description of the line and reflection features allows less
freedom to fit the spectrum than the phenomenological
components, and shows that there is more spectral complexity in
the data (predominantly in S7) than is described by reflection.


\begin{table*}
\begin{tabular}{lccccccccc}
\hline
Obs. & $kT_{\rm disc}$ & $\Gamma$ & $kT_{\rm comp}$ & $\Omega/2\pi$ & $\log(\xi)$ & $R_{\rm in}$ & $L_{\rm disc}$ & $L_{\rm comp}$ & $\chi^2/\nu$ \\
& (keV) & & (keV) & & (erg cm s$^{-1}$) & ($R_g$)  & \multicolumn{2}{c}{($10^{38}$ erg s$^{-1}$)}\\
\hline

1 & $<0.62$ & $2.07_{-0.04}^{+0.02}$ & $>18$ & $0.15\pm
0.04$ & $3.7_{-0.5}^{+0.2}$ & $43_{-20}^{+80}$ & $ <1.8$ & $5.3_{-0.6}^{+0.9}$ & 31.1/37\\
2 & $0.91\pm 0.09$ & $2.30\pm 0.02$ & $> 17$ & $0.17_{-0.04}^{+0.14}$ & $4.0_{-0.7}^{+0.6}$ & $43_{-23}^{+75}$ & $3.36_{-0.13}^{+0.15}$ & $4.0_{-0.2}^{+0.4}$ & 28.9/37\\
3 & $1.96\pm 0.25$ & $1.61_{-0.47}^{+0.23}$ & $4.2_{-0.4}^{+0.6}$ &
$ > 1.26$ & $<2.6 $ & $41_{-14}^{+40} $ & $5.2_{-1.1}^{+0.4}$ & $1.7_{-0.3}^{+1.3}$ & 21.0/37\\
4 & $0.80_{-0.24}^{+0.31}$ & $2.35_{-0.16}^{+0.05}$ & $6.4\pm0.7$ &
$0.18_{-0.02}^{+0.04}$ & $3.4_{-0.4}^{+0.3}$ & $> 100$ &
$<4.5 $ & $5.6_{-1.5}^{+3.0}$ & 46.6/37\\
5 & $0.79_{-0.31}^{+0.12}$ & $2.48_{-0.04}^{+0.08}$ & $>6.2 $
& $0.18_{-0.02}^{+0.04}$ & $3.4^{+0.3}_{-1.4}$ & $ >100 $ &
$<5.6 $ & $5.6_{-0.7}^{+3.3}$ & 34.5/37\\
6 & $0.94_{-0.02}^{+0.04}$ & $3.03_{-0.07}^{+0.10}$ &
$3.53_{-0.17}^{+0.11}$ & $0.50_{-0.11}^{+0.16}$ & $<1.9$ &
$40_{-24}^{+230}$ & $<1.9 $ & $10.7_{-0.9}^{+1.0}$ & 16.8/37\\
7 & $1.17_{-0.06}^{+0.03}$ & $1.5^{+0.5}_{-0.3}$ &
$2.58^{+0.30}_{-0.15}$ & $1.40_{-0.22}^{+0.19}$ & $3.6_{-0.2}^{+0.1}$
& $>350 $ & $6.8_{-0.3}^{+0.2}$ & $0.5_{-0.1}^{+0.3}$ & 208/37\\

\hline

\end{tabular}

\caption{Best-fitting parameters for the selected spectra using
a model with relativistically smeared, ionized reflection (M2 in
Table~\ref{tab:models}).}

\label{tab:reflection}
\end{table*}


Fig.~\ref{fig:residuals} (panel M2) shows the residuals for the
reflection model of S7, where the main mismatch is a marked excess in
the data at $\sim$5.5~keV. This feature is systematically present in
the other 6 spectra also, but is much less significant. While such
features are obviously indicative of a strongly redshifted, broad
shoulder to the iron line emission, these effects are already included
in our model spectra (which has iron at 3$\times$ solar abundance, so the
line is extremely strong).  Instead this most probably represents the
effects of excess {\em absorption}.  Individual edges from neutral
elements can be resolved with the {\it Chandra\/} grating spectra, and
these clearly show that the neutral absorption has strong abundance
anomalies, with Si and Fe being a factor 3.5 and 2.3 overabundant,
respectively (Lee et al.~2002). This most probably arises from
material ejected during the formation of the black hole (Lee et
al.~2002), so it should be fairly far from the system, and constant
with time.  We use the {\sc varabs} code to model this, fixing the
column of all the elements at $4.7\times 10^{22}$ cm$^{-2}$ except for
Si and Fe which are fixed at $16.4$ and $10.9\times 10^{22}$ cm$^{-2}$
respectively as measured by {\it Chandra} (Lee et al.~2002). We call
this model M3 (Table \ref{tab:models}).

We do not tabulate the detailed results of this intermediate
model and only note that the increased absorption improves the
fit for no increase in the number of degrees of freedom for all
the data except S3. While $\chi^2/\nu$ = 34/37 for S3 is
somewhat worse than that for the standard abundance model, it is
an adequate fit to the data. However, S7 still has an
unacceptable $\chi^2/\nu= 93/37$, as shown by its residuals in
Fig.~\ref{fig:residuals} (panel M3).

High- and moderate-resolution spectral data indicate yet more
complexity in the absorption, in that as well as the anomalous
abundance neutral absorption there is clear evidence for ionized
absorption.  Both {\it ASCA\/} and {\it Chandra\/} detect resonance
line absorption from highly ionized iron, and show that this is {\em
variable} with time and/or spectral state of \grs\ (Kotani et
al. 2000; Lee et al.~2002).  We model this by narrow Gaussian lines
corresponding to He- and H-like iron, with K$\alpha$ transitions at
6.67 and 6.9~keV, respectively. We also include the (fixed energy)
K$\beta$ and $\gamma$ lines at $7.80, 8.21$~keV for He-like and $8.16,
8.61$~keV for H-like, with fixed relative intensity to the K$\alpha$
transition of 0.07 and 0.06 for He-like and 0.2 and 0.07 for H-like
(Kotani et al. 2000). Thus there are only two additional free
parameters, which are the intensities of the 6.67 and 6.90~keV
absorption lines, denoted as $I_{6.67}$ and $I_{6.90}$,
respectively. These are not strongly detected in S1--6, but the fit to
S7 is significantly improved, and now gives an acceptable
$\chi^2/\nu=34.2/35$. Residuals to this fit are shown in
Fig.~\ref{fig:residuals} (panel M4).

We use the {\sc xstar} code (Kallman et al.~1996) to check the
assumption that the dominant opacity is from the iron absorption
lines (as opposed to lines from other elements and/or
photoelectric edges). We assume the same Si/Fe abundance
anomalies as in the neutral absorption, illuminated with a
$\Gamma=2$ power law. This self-consistent absorption model
gives an identical $\chi^2$ to that of the resonance absorption
lines model of Table~\ref{tab:absorption}.  The resulting
parameters imply these lines arise in an equivalent hydrogen
column of $\sim 3\times 10^{23}$ cm$^{-2}$ with ionization
parameter $\xi \sim 10^{3.7}$ ergs cm s$^{-1}$.  None of the
seven spectra are significantly better fit with the full {\sc
xstar} ionized absorption models so we use the absorption lines
in all subsequent fits as they are easier to constrain.

Our final model M4 consists of the disc emission, its thermal
Comptonization, ionized and relativistically smeared reflection,
cold absorption with anomalous abundance of silicon and iron and
resonance absorption lines from highly ionized iron. The fit
results to all the seven spectra are in
Table~\ref{tab:absorption}.

The effective increase in cold absorption affected mostly the
soft part of the spectrum, decreasing the soft colour by
$\sim$0.2, when M2 was replaced by M4. The extrapolated
bolometric disc luminosity is also heavily dependent on the
assumed absorption model. With model M4 the disc luminosity is
significantly larger than with M2. This is particularly well
seen for spectrum S3, where changing the absorption model has
switched the fit from a high disc temperature ($\approx$ 2~keV),
low Comptonized luminosity model, to a low disc temperature
($\approx$ 0.7~keV), high Comptonized luminosity model. In S6
the best-fitting M4 model is disc-dominated ($\sim$0.1 of total
power in the Comptonized tail), though this particular fit is
very poorly constrained and a more Comptonized spectrum
($\sim$0.6 luminosity in the tail) is also possible. This shows
the difficulties in dealing with such an absorbed source, and
illustrates the reason for the conservative limit of DG03 of
restricting their sample to objects with column of $< 2\times
10^{22}$ cm$^{-2}$.

We note that none of the models including reflection (M2--4) give any
indication for the extreme relativistically smeared components
expected for an X-ray illuminated disc which extends down to the last
stable orbit around a black hole. All the inner radii derived from
relativistic smearing are substantially larger than $\sim$20~$R_g$
(10 Schwarzchild radii). Similarly large radii are also inferred by
Martocchia et al. (2002) from {\it BeppoSAX\/} data on this source.
This is important as there is some evidence that \grs\ is an extreme
Kerr black hole (Zhang et al. 1997; Cui et al. 1998; Sobolewska \&
{\.Z}ycki 2003), with the disc extending down to 1.24~$R_g$. However,
the line broadening does not rule out such a disc since the reflector
is highly ionized. The innermost disc may well be so highly ionized
that iron in the disc is completely stripped, so there are no line
features produced in the most strongly curved space-time regions
though there can be continuum emission/reflection from the disc (see
also Wilson \& Done 2001 for a similar conclusion for the very high
state spectra of XTE~J1550--564).


\begin{table*}
\begin{tabular}{lccccccccccc}
\hline

Obs. & $kT_{\rm disc}$ & $\Gamma$ & $kT_{\rm comp}$ & $\Omega/2\pi$ & $\log(\xi)$ & $R_{\rm in}$ & $I_{6.67}$ & $I_{6.95}$ & $L_{\rm disc}$ & $L_{\rm comp}$ & $\chi^2/\nu$ \\
& (keV) & & (keV) & & (erg cm s$^{-1}$) & ($R_g$) & \multicolumn{2}{c}{($10^{-3}$ cm$^{-2}$ s$^{-1}$)} & \multicolumn{2}{c}{($10^{38}$ erg s$^{-1}$)}\\
\hline

1 & $<0.45 $ & $2.07\pm 0.03$ & $>12 $ & $0.20\pm 0.03$ & $3.8_{-0.5}^{+0.6}$ & $11\pm 3$ & $<1.6 $ & $<3.7 $ & $10.9_{-3.6}^{+1.6}$ & $4.9_{-0.4}^{+1.1}$ & 26.4/35 \\
2 & $0.55^{+0.11}_{-0.10}$ & $2.37^{+0.02}_{-0.06}$ & $>25 $ & $0.34_{-0.08}^{+0.24}$ & $4.3\pm0.5$ & $18_{-4}^{+10}$ & $<3.0 $ & $7.1^{+5.1}_{-2.3}$ & $6.8_{-0.6}^{+2.6}$ & $5.0_{-0.7}^{+1.3}$ & 15.2/35\\
3 & $0.70_{-0.16}^{+0.17}$ & $2.11\pm 0.05$ & $5.54_{-0.37}^{+0.44}$ & $0.23^{+0.07}_{-0.03}$ & $3.3\pm0.4$ & $25_{-7}^{+15}$ & $< 3.5 $ & $<8.3 $ & $<2.2 $ & $6.0_{-0.1}^{+0.3}$ & 33.7/35\\
4 & $<0.47 $ & $2.30\pm 0.04$ & $5.53_{-0.33}^{+0.41}$ & $0.25_{-0.04}^{+0.06}$ & $3.5_{-0.5}^{+0.4}$ & $34_{-12}^{+30}$ & $<5.3 $ & $<12.4 $ & $12.5_{-5.9}^{+5.7}$ & $8.7_{-0.3}^{+0.7}$ & 38.3/35\\
5 & $<0.47 $ & $2.51_{-0.05}^{+0.04}$ & $6.55_{-0.53}^{+0.63}$ & $0.23\pm 0.06$ & $3.5\pm0.5$ & $45_{-19}^{+71}$ & $<7.7 $ & $<13.1 $ & $20.9_{-7.7}^{+6.9}$ & $10.2_{-0.3}^{+0.8}$ & 18.9/35\\
6 & $1.61_{-0.61}^{+0.08}$ & $<2.85 $ & $2.78_{-0.08}^{+0.52}$ & $>0.12 $ & $4.9_{-2.7}^{+0.3}$ & $ * $ & $ *$ & $ <18 $ & $16.2_{-9.8}^{+0.9}$ & $1.2_{-0.5}^{+8.0}$ & 9.8/35\\
7 & $0.99_{-0.05}^{+0.07}$ & $1.93_{-0.20}^{+0.41}$ & $2.43_{-0.08}^{+0.14}$ & $1.88_{-0.38}^{+0.12}$ & $3.3\pm0.3$ & $>150$ & $22^{+9}_{-13}$ & $14^{+7}_{-6}$ & $7.9_{-0.4}^{+0.3}$ & $0.9_{-0.2}^{+0.6}$ & 34.2/35\\
\hline
\end{tabular}

\caption{Best-fitting parameters for the selected spectra
(Table~\ref{tab:Obs IDs}) using our best physical model with
reflection, anomalous neutral abundances and ionized iron
absorption lines (M4 in Table~\ref{tab:models}).
An asterisk indicates that the parameter is
completely unconstrained. }

\label{tab:absorption}
\end{table*}


\subsection{Physical model applied to all pointed spectra}

\begin{figure*}
\centerline{{\psfig{file=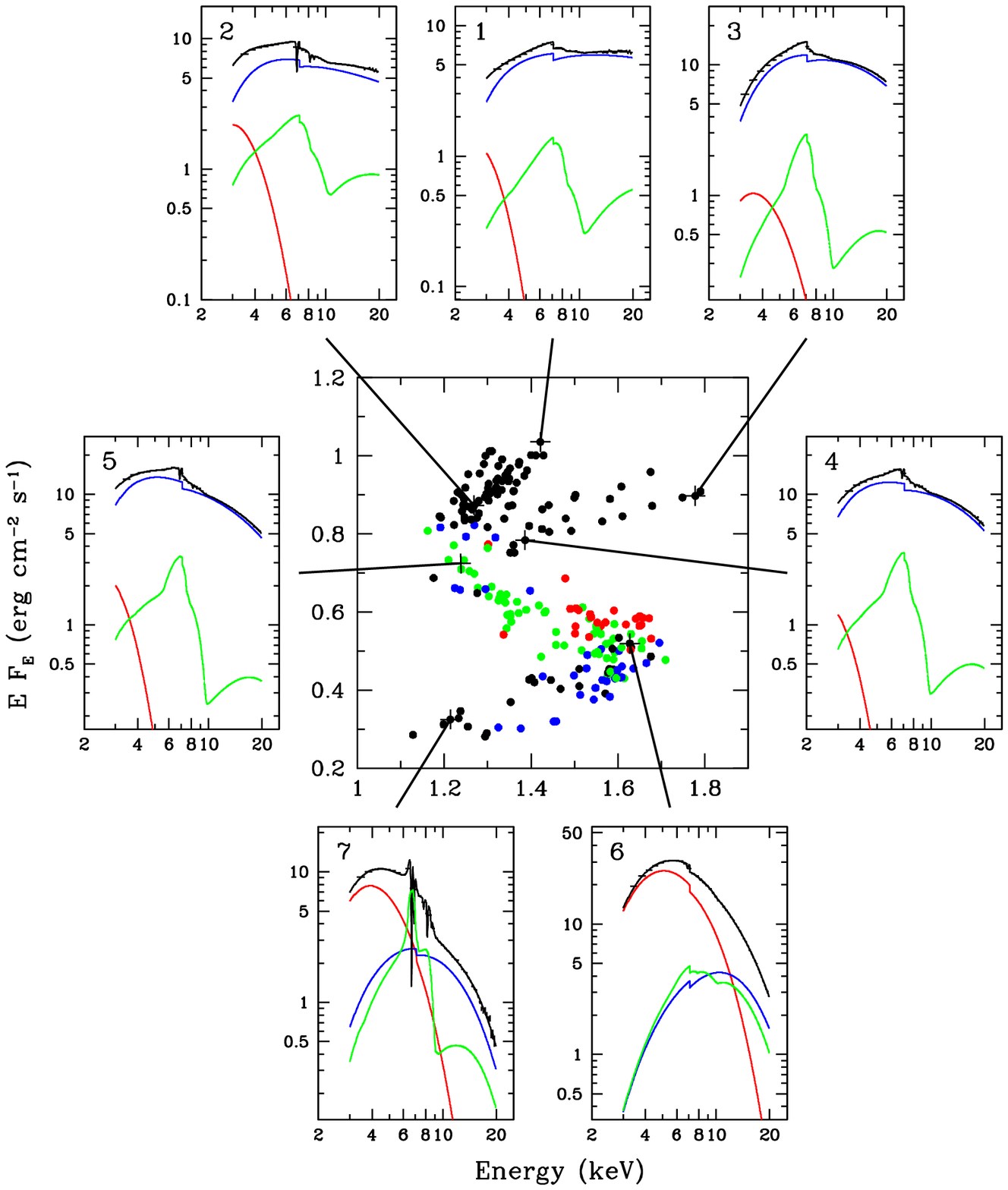,width=0.8\textwidth}}}

\caption{The intrinsic colour-colour plot for all the
observations shown in Fig.~\ref{fig:rms} fitted with the
physical model including reflection, anomalous neutral
abundances and ionized iron absorption lines (M4, Table
\ref{tab:models}). The individual spectra corresponding to the
selected observations listed in Table \ref{tab:Obs IDs} are shown
deconvolved with this model, where the red, blue and green lines
give the disc, Comptonized emission and its reflection,
respectively. The best-fitting parameters for each spectrum are
shown in Table \ref{tab:absorption}. }

\label{fig:spider}
\end{figure*}

The previous section has shown that a physically motivated model
for \grs\ which can adequately describe {\em all} the selected
spectra includes anomalous abundance neutral absorption, ionized
absorption, relativistically smeared, ionized reflection, with
continuum from the disc and thermal Comptonization.  The high
absorption means that the derived soft colour and disc
luminosity change substantially with changing the model.
Nonetheless, we now have a physical model which incorporates all
the observed spectral features so we can at least use this to
get our best estimate for the intrinsic colours of the system.
We refit the 292 spectra from Fig.~\ref{fig:rms} with this
physical model, and the resulting colour-colour diagram is shown
in Fig.~\ref{fig:spider}. Black and blue points on this diagram
correspond to spectra which show little variability, so the model
parameters represent the physical components of the spectra.
However, this is not the case for the spectra averaged over
high-variability periods (shown in green and red in
Figs.~\ref{fig:rms} and \ref{fig:spider}), where the physical model
is merely parameterizing the time averaged spectrum to give a
best estimate for the colours.  A comparison with
Fig.~\ref{fig:rms} confirms the result from the individual
spectral fits that the major change is that the soft colour is
reduced by $\sim$0.2 in all spectra, so that the overall
pattern stays the same.

Fig.~\ref{fig:spider} also shows the best-fitting unfolded
individual spectra, together with the model components. Spectra
along the upper diagonal branch between S1 and S2 have a hard
component which has no discernable rollover in the PCA bandpass
($kT_{\rm comp} \ga 10$~keV), with spectral index steepening
from S1 to S2. S3--S5 form a parallel diagonal track for
changing spectral index, but here the electron temperature is
clearly seen in the high-energy spectral curvature ($kT_{\rm
comp} \sim 6$~keV). All these spectra have comparable disc and
hard tail luminosity. S6 and S7 are different from S1--S5 in
that there is an increase in disc temperature. The hard tail is
characterized by even lower temperature electrons, $kT_{\rm
comp} \sim$ 2--3~keV. The disc temperature $\ga$1~keV rather
than hard tail dominates the change in soft colour.

Only spectra with colours similar to those of S7 show a strong
detection of the resonance absorption lines from He- and H-like Fe.
This is consistent with previous moderate- and high-resolution
spectra from {\it ASCA\/} and {\it Chandra}. The {\it Chandra\/}
spectrum of Lee et al. (2002) has a continuum similar to S2, with
resonant line intensities of a few $10^{-3}$ photons cm$^{-2}$
s$^{-1}$, easily consistent with our upper limit in
Table~\ref{tab:absorption}. The three {\it ASCA\/} spectra of Kotani
et al. (2000) taken in 1994 September, 1995 April and 1996 October,
have colours similar to S7, midway between S6 and S7, and similar to
S5, respectively. The lines are only significantly detected in the
1994 and 1995 {\it ASCA\/} data, where they have equivalent widths of
30--50~eV for the He- and H-like K$\alpha$ lines. This is somewhat
smaller than, but comparable to the 100--120~eV equivalent widths for
these lines we derive for S7. This material is almost certainly part
of an outflowing wind, which can have a comparable mass loss rate to
that of the mass accretion rate required to power the observed
emission (Lee et al.~2002). That the lines are most visible in S7 can
be explained as an ionization effect. All the other spectra are
harder, so a higher proportion of iron can be completely ionized,
giving smaller equivalent width features. This can also feedback onto
the wind acceleration mechanism if there is considerable line
driving, so that there is intrinsically less wind at higher
ionization (e.g. Proga \& Kallman 2002).


\section{\grs\ in context}


\begin{figure*}
{\psfig{file=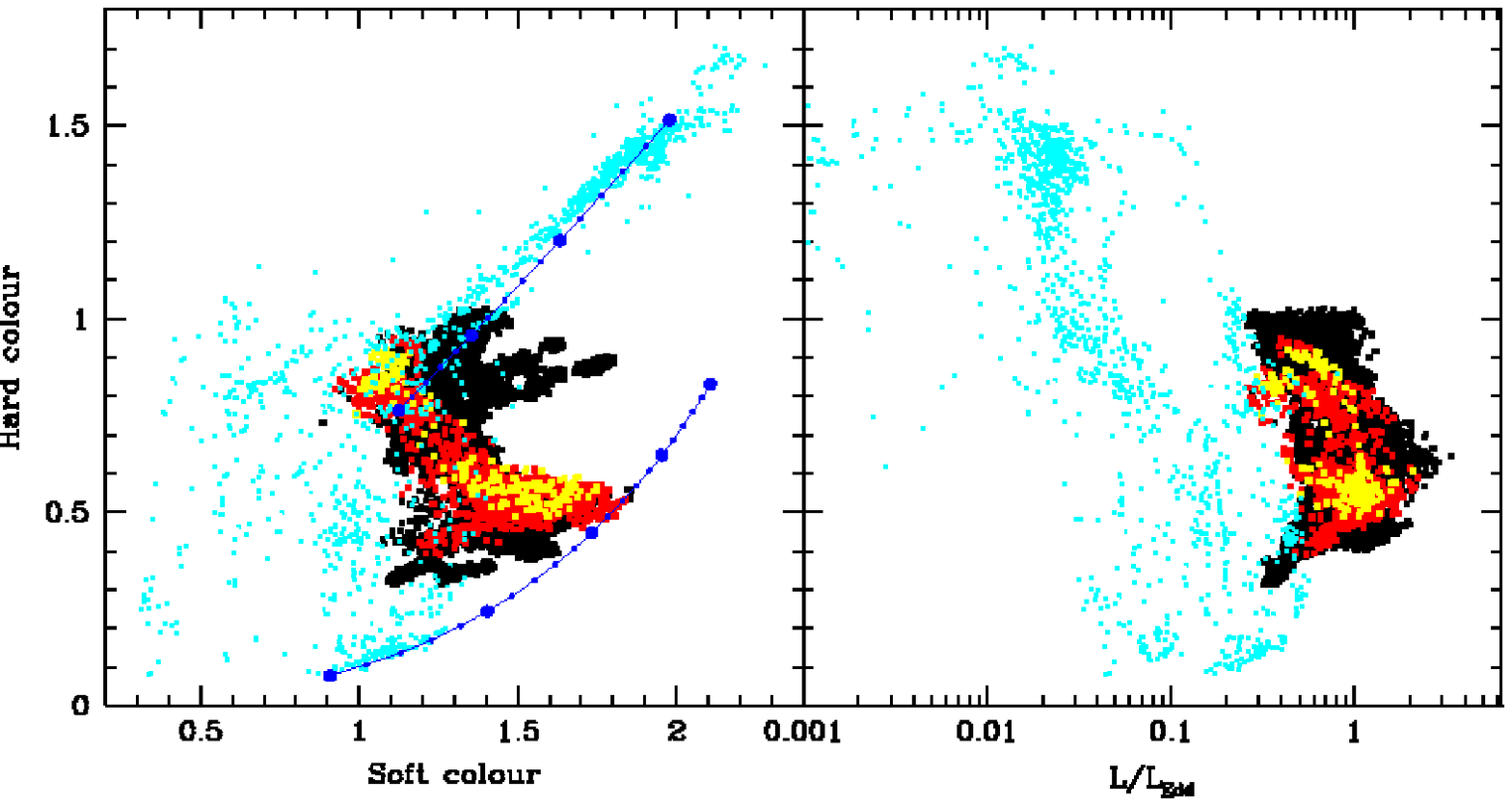,width=0.9\textwidth}}

\caption{The intrinsic colour-colour (left) and colour-luminosity
(right) diagrams for \grs\ created from 128-s spectra. Black, red and
yellow points denote periods of low, high and very high variability
with r.m.s. $<$ 40, 40--60 and $\ge$ 60 per cent, respectively. On
top of this we have superimposed the data from `normal' black holes (cyan
points) from DG03 and GD03. The blue lines on the left diagram show
the model colours of a power law (upper line, with $\Gamma$
increasing from 1.5 to 3.0 diagonally downwards) and a multicolour disc
(lower line, with $kT_{\rm disc}$ increasing from 1 to 3 keV
rightwards). The yellow points track the highest variability of \grs,
related to limit-cycle instability. It occurs between colours
corresponding to the very
high state (1.1,~0.8) and a hot ultrasoft state
corresponding to an effective disc temperature of excess of 2~keV.}

\label{fig:colum}
\end{figure*}


While our physically based models give generally good fits to all
the individual observation ID spectra, these data are extracted
from full {\it RXTE\/} pointings. These have exposure times of
several kiloseconds, so many of them average over periods of
strong spectral variability. Thus the parameters derived from the
spectral fits in Fig.~\ref{fig:spider} will not show the extremes
of luminosity, and can be misleading since the model components
are not linear.

Thus we need to go back to the 128-s spectra in order to get a
clearer idea of the properties of \grs. However, it is not
feasible to fit all 16489 spectra with the computationally
intensive physical model, where a single fit can take tens of
minutes. Fortunately, the spectral fits from
Section~\ref{sec:ObsIDs} show that most of the change in derived
soft colour and soft luminosity between the simple
phenomenological description and the more physical model arises
from the change in cold absorption. Yet most of the increase in
fitting time comes from modelling reflection. Hence we use the
simple phenomenological model of smeared edge and broad line to
describe reflection (as in model M1 of Table~\ref{tab:models}),
with fixed anomalous abundances of Si and Fe to test how well
this can reproduce the colours and luminosities derived from the
physical model (we call this model M5; see
Table~\ref{tab:models}). A fit of this simple model to the seven
selected spectra from Section~\ref{sec:ObsIDs} shows that it
gives a very good match to the colours from the physical model,
and reproduces the soft luminosities to within a factor of
$\sim$2. This gives some measure of the systematic uncertainty
in deriving the total (model dependent) luminosity.

We use this simplified model (M5) to fit all the 128-s spectra. The
left panel in Fig.~\ref{fig:colum} shows the colour-colour plot
derived from this (black, red and yellow points), compared to the
colours derived from the `normal' black holes of DG03 and GD03 (cyan
points). DG03 show how position on this diagram relates to the
different black hole spectral states (see e.g. the reviews by Tanaka
\& Lewin 1995; Esin, Mclintock \& Narayan 1997), which we briefly
summarize here. The power law dominated {\em low/hard state\/}
corresponds to the diagonal track starting from colours (2.0,~1.5)
for an photon spectral index of $\Gamma \sim 1.5$, extending down to
about (1.5,~1.0) as the spectrum softens to $\Gamma \sim 2$. {\em
Intermediate\/} or {\it very high state spectra\/} have roughly equal
luminosity in a disc component as in a fairly steep ($\Gamma \sim$
2--2.5) power-law tail and have colours within $\sim$0.2 of
(1.3,~0.8). The disc is completely dominant in the {\em ultrasoft
state}, so the colours are close to those predicted by the pure disc
blackbody track. The classic {\em high/soft state} (as seen e.g. in
Cyg X-1: Gierli{\'n}ski et al. 1999) has a disc-dominated soft
spectrum but at higher energies has a noticeable fraction of
luminosity in a power-law tail. This gives it somewhat harder colours
than the ultrasoft states at $\sim$(0.7,~0.8). This is the region
identified by DG03 as a sufficient (but not necessary) condition for
the object to be a black hole.

The spectra from \grs\ with colours similar to S1--S5 overlap with
the intermediate/very high state region on the colour-colour diagram,
while S6 and S7 occupy the high-temperature end of the ultrasoft
region. \grs\ {\em never\/} goes into the classic low/hard state
$\sim$(2.0, 1.5), nor does it go into the `black hole only' region
of DG03 (the classic high/soft state).

We categorize each 128-s spectra according to variability of a period
they come from, marking them in black, red and yellow for r.m.s. $<$
40, 40--60 and $\ge$ 60 per cent, respectively. This makes it clear
that the major instability track is between spectra with colours like S2 and
S6 (see Fig.~\ref{fig:rms}). A comparison with Fig. \ref{fig:spider}
shows that the higher time resolution of these spectra (128-s as
opposed to a few ks) gives a similar diagonal track for
the limit cycle, but extending further towards the disc blackbody line
(i.e. corresponding to higher temperatures and luminosities).
With even higher resolution data then this track extends even further
(see the solid and dashed line in Fig.~\ref{fig:bell_col}a), showing that
our 128-s spectra do not follow the fastest variability of \grs.

The upper-left end of the track corresponds to spectral state C
(Fig.~\ref{fig:bell_col}b), rather misleadingly also called the hard
state. In fact, it is very similar to e.g. very high state data from
XTE J1550--564 (Gierli{\'n}ski \& Done 2003a). There is nothing
special in the spectral shape here which hints at the origin of the
unique variability of \grs. The other end of the instability track
lays in the ultrasoft region, corresponding predominantly to the
higher temperature state B (Fig.~\ref{fig:bell_col}b). These have
similar colours to pure disc spectra, but are at a higher effective
temperature than any of the other black holes. Thus the only hint of
difference in the colours for the origin of the dramatic instability
in \grs\ is that the ultrasoft state appears to extend to higher disc
temperatures.

The right panel of Fig.~\ref{fig:colum} is a colour-luminosity plot
for \grs\ superimposed on that for all the other black holes (DG03,
GD03). The colour-coding is the same as in the left panel. This is
much more illuminating as to the origin of the unique instability
behaviour of \grs.  Here it is plain  that the instability occurs in
the most luminous spectra, those which have $L > L_{\rm Edd}$, and
{\em only\/} \grs\ goes up to such high luminosities. The luminosity
never drops below $\sim$0.3 $L_{\rm Edd}$, explaining why \grs\ never
goes into the hard state (typically seen only at luminosities lower
than a few percent of Eddington). It also explains why \grs\ {\em never\/}
goes into the `black hole only' area of the colour-colour diagram.
Spectra in this region, characterized by a low-temperature disc
component together with a weak hard tail, are seen predominantly
around $\sim$0.1 $L_{\rm Edd}$. The persistently high luminosity of
\grs\ takes it above these low disc temperatures.

Where \grs\ has a luminosity which overlaps with that of the `normal'
black holes it shows comparable spectra and does not show the dramatic
limit cycle variability.  However, it also extends to higher
luminosities, up to $\sim$2--3 $L_{\rm Edd}$. This makes it unique
among the black holes studied by DG03 and GD03, and it is here that
the unique instability occurs. However, supereddington luminosity
seems to be only a necessary, not sufficient condition for the disc
instability. There are several low variability spectra from \grs\
which exceed the Eddington limit. Perhaps the instability requires
that the {\em disc} as opposed to total luminosity goes above
Eddington, or perhaps this merely shows that the disc structure takes
some time to respond to an increase in central mass accretion rate
(e.g. van der Klis 2001).

The left panel in Fig.~\ref{fig:colum} includes the background colour
tracks for a disc blackbody spectrum, and the high luminosity,
ultrasoft spectra on the extreme end of the limit-cycle variability
intersect this line at colours corresponding to $kT_{\rm disc}\sim
2.2$~keV. A more detailed analysis is complicated as many of these
most extreme colour 128-s spectra still average over substantial
variability. However, these ultrasoft spectra have similar colours to
the stable spectrum S6. {\em If\/} they are truly similar then the
results in Table~\ref{tab:absorption} indicate that these are not
simply high-temperature disc emission. Instead they might be much
better fit by a lower temperature disc at $\la$1.6~keV, with a
fairly strong ($\ga$10 per cent of bolometric flux), low temperature
(few keV) Comptonized tail (see also Zdziarski et al.~2001). A
similar effect is also seen in the high-temperature ultrasoft spectra
of GRO~J1655--40 (Kubota, Makishima \& Ebisawa 2001; GD03).

However, irrespective of the detailed spectral form of the highest
luminosity spectra, we include the {\em minimum\/} absorption column,
so derive (on average) a lower limit to the flux. Additional, time
variable cold absorption components (e.g. Sobolewska \& {\.Z}ycki
2003) will only {\em increase\/} the inferred luminosity.  There is a
factor $\sim$2 systematic uncertainty on the distance (7--12 kpc:
Fender et al. 1999, though the lower limit could present significant
problems for the derived luminosity of the companion star: Greiner et
al 2001b) so this could reduce the bolometric luminosity estimates by
at most a factor 4.  Thus it seems that the trigger for the
characteristic limit-cycle variability in \grs\ is almost certainly
super-Eddington accretion rates.

\section{Limit-cycle variability}

\grs\ is the most luminous black hole system studied here, and is
also the only one to show the characteristic limit-cycle variability.
Plainly the results are consistent with the accretion flow onto a
black hole becoming unstable at very high luminosities of
$\ga L_{\rm Edd}$. The same instability does not seem to operate in
neutron stars. Z sources emit at 1--3 $L_{\rm Edd}$, while
Cir X-1 can reach up to $10 L_{\rm Edd}$ (DG03). Approximately half
of the neutron star luminosity should be from the boundary layer, so
the accretion flow luminosity is only $\sim$0.5--1.5 $L_{\rm Edd}$ in
the Z sources, perhaps too low to trigger the instability (but not in
Cir X-1). Alternatively, irradiation of the accretion flow by the
boundary layer may act as a stabilizing mechanism (Czerny, Czerny \&
Grindlay 1986, but see Mineshige \& Kusenose 1993 and below).

Following Belloni et al. (1997a, b), the instability itself is
generally associated with the well known radiation pressure
instability of a standard $\alpha$ disc at high luminosities
(Lightman \& Eardley 1974; Shakura \& Sunyaev 1976). In its simplest
form this predicts that the inner accretion flow cycles between a
highly luminous, hot, advection-dominated disc and a much lower
luminosity, cooler, gas pressure-dominated disc (Abramowicz et al.
1988; Honma, Matsumoto \& Kato 1991; Szuszkiewicz \& Miller 1997,
1998; Zampieri, Turolla \& Szuszkiewicz 2001). One key problem with
this interpretation is that the limit-cycle variability should occur
at well below $L_{\rm Edd}$ (see Nayakshin, Rappaport \& Melia 2000
for other problems). Theoretical models show that the limit cycles
become noticeable at $L_{\rm disc} \ga 0.3L_{\rm Edd}$ (Janiuk, Czerny
\& Siemiginowska 2002), yet {\em none} of the comparison sample of
`normal' black holes above this luminosity show the limit cycle
(Kubota et al. 2001; Kubota \& Makishima 2003; GD03).

GD03 summarize various ways to delay the onset of the radiation
pressure instability in a standard disc. Generally these involve
some additional energy loss channel e.g. dissipating some of the
accretion energy in a hard X-ray corona, or via a wind or jet.
The problem with these is that the `normal' black holes can show
disc-dominated spectra (so no obvious energy loss in a
corona/jet) without producing the instability even at $L_{\rm
disc} \sim 0.5 L_{\rm Edd}$ (GD03). By contrast \grs\ shows
strong coronal emission {\em and} has a jet {\em and} a wind,
{\em all} of which should act to stabilize the disc, yet {\em
does} show the instability. Irradiation of the disc by the
strong coronal emission should also act as a stabilizing
mechanism, similarly to that proposed for the boundary layer in
neutron stars (Czerny et al. 1986).  Thus it seems unlikely that
irradiation alone can stabilize the disc in the Z sources since
it does not manage to stabilize \grs.

The `normal' black holes show that the disc is stable up to at least
$L_{\rm disc} \sim 0.5L_{\rm Edd}$, so its structure must be rather
different from that predicted by the standard $\alpha$ disc equations
(GD03). The fact that the instability in \grs\ is triggered at or
near Eddington luminosity seems to suggest that radiation pressure
does play a pivotal role in the formation of the dramatic
variability. That the $\alpha$ disc equations do not work in detail
is perhaps not surprising given the {\it ad hoc} nature of viscous
heating in these models. The next generation of disc models will give
a much better estimate for the disc structure, as these can
incorporate the MHD dynamo which is the physical origin of the
viscosity (Balbus \& Hawley 1991). When these models are sufficiently
developed to solve the time dependent 3D, coupled radiation and
magneto-hydrodynamic equations over a substantial range of radii
(Turner, Stone \& Sano 2002; Turner et al.~2003) then the question of
inner disc stability for highly luminous systems can be revisited.

\section{Long-term variability}

The unique short-term limit-cycle variability of \grs\ is plausibly
because it reaches higher $L/L_{\rm Edd}$ than any other known
Galactic black hole. But then the question becomes why \grs\ alone
reaches such high luminosities? The answer to this lies in the
evolutionary state of \grs. The system has the longest known period
of any LMXB of 33.5 days, implying a very wide orbit (Greiner et al.
2001a). The secondary is a low-mass giant star (Greiner et al.
2001b), so the accretion takes place via Roche lobe overflow
(Eikenberry \& Bandyopadhyay 2000).

There is now considerable agreement that transient outbursts in the
LMXBs are caused by the classic disc instability mechanism which
operates when hydrogen goes from being mainly neutral to mainly
ionized.  This is a very different physical mechanism for a disc
instability than the radiation pressure instability discussed above,
and happens at much lower luminosity, but it gives rise to another
S-shaped thermal equilibrium curve (Smak 1982). In discs around white
dwarfs it can trigger limit cycles where there are disc outbursts on
timescales of days--weeks, followed by a period of quiescence lasting
weeks--months (see e.g. the review by Osaki 1996). In LMXBs the disc
structure is strongly modified by irradiation once the outburst
starts, keeping hydrogen ionized for much longer so that the outburst
timescales are dramatically extended (van Paradijs 1996; King 1997;
King \& Ritter 1998; Dubus et al.~2001).

The peak luminosity at the start of the LMXB outburst is determined
by the disc size, $R_T$. The maximum mass the disc can hold before
going into outburst is $\propto R_T^3$, while the inflow timescale is
$\propto R_T$. Thus the peak luminosity $\propto R^2_T$ so the bigger
the disc, the bigger the peak luminosity (King \& Ritter 1998). But
the disc size in LMXB is fixed by tidal torques to $\sim$1.4$\times
R_{\rm circ}$, where $R_{\rm circ}$ is the circularization radius of
the binary (Shahbaz, Charles \& King 1998), and the binary system
parameters are determined by the requirement that the
secondary star can overflow its Roche lobe. For main sequence
secondaries the orbit must be fairly small, so the disc is small.
\grs\ has a giant secondary, so has a huge disc, with $R_T\sim
10^{12}$ cm for a mass ratio of $0.1$ (e.g. Frank, King \& Raine
1992). The quiescent disc mass is $\sim$10$^{28}$ g (Shahbaz, Charles
\& King 1998), which can maintain an Eddington accretion rate for
$\sim$10 years, similar to observed outburst timescale.

Thus both the long and short-term unique variability behaviour
of \grs\ can be explained as a function of its evolutionary
state, where the wide orbit allows a huge disc to form.

\section{Conclusions}

\grs\ is not in a separate class from `normal' black holes. When
it is at the same $L/L_{\rm Edd}$ then its spectra and (more
importantly) time variability behaviour are similar to that seen
in the 'normal' black holes. Its unique limit cycle variability
only appears when it radiates at uniquely high (super Eddington)
luminosities.

\grs\ boldly goes to luminosities where no black hole has gone before, and
we eagerly await the next generation of disc models which will more clearly
identify the origin of the instability.


\section*{Acknowledgements}

We thank Ed Cackett for stimulating discussions. GW support from
the Royal Society, KBN grant 5P03D00821 and the Foundation for
Polish Science.

\end{document}